\documentclass{article}

\usepackage{arxiv}

\usepackage[utf8]{inputenc} 
\usepackage[T1]{fontenc}    
\usepackage{hyperref}       
\usepackage{url}            
\usepackage{booktabs}       
\usepackage{amsfonts}       
\usepackage{nicefrac}       
\usepackage{microtype}      
\usepackage{lipsum}     
\usepackage{graphicx}
\usepackage{doi}

\usepackage{amsmath}
\usepackage{amssymb}
\usepackage{latexsym}
\usepackage{graphicx}
\usepackage{natbib}
\usepackage{color}
\bibpunct{(}{)}{;}{a}{}{,}

\usepackage{amstext}

\usepackage{booktabs}
\usepackage{stackengine}
\usepackage{nccmath}

\newcommand{\apj}{ApJ}
\newcommand{\aap}{A\&A}
\newcommand{\apjl}{ApJL}
\newcommand{\apjs}{ApJS}
\newcommand{\nat}{Nature}
\newcommand{\solphys}{Sol. Phys.}
\newcommand{\ssr}{Space Sci. Rev.}
\newcommand{\araa}{ARA\&A}
\newcommand{\mnras}{Monthly Notices of the Royal Astronomical Society}

\title{On the origins of coronal Alfv\'enic waves}

\author{ \href{https://orcid.org/0000-0001-5678-9002}{\includegraphics[scale=0.06]{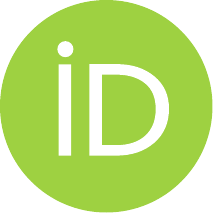}\hspace{1mm}Richard J. Morton}\\
Department of Maths Physics and Electrical Engineering \\
Northumbria University, UK \\
\texttt{richard.morton@northumbria.ac.uk}\\
\And
\href{https://orcid.org/0000-0001-6121-7375}{\includegraphics[scale=0.06]{orcid.pdf}\hspace{1mm}Roberto Soler}\\
Departament de F\'{i}sica \\ Universitat de les Illes Balears\\ E-07122, Palma de Mallorca \\ Spain \\
}

\begin{document}
\maketitle

\begin{abstract}
Alfv\'enic waves are considered a key contributor to the energy flux that powers the Sun's corona, with theoretical models demonstrating their potential to explain coronal EUV and X-ray emission and the acceleration of the solar wind. However, confirming underlying assumptions of the models has proved challenging, especially obtaining evidence for the excitation and dissipation of Alfv\'enic waves in the lower solar atmosphere and tracing their propagation into the corona. 
We present an investigation of the Alfv\'enic wave power spectrum in the Sun’s corona, obtained from observations with DKIST Cryo-NIRSP. The data provide unprecedented temporal resolution and signal-to-noise, revealing a detailed power spectrum out to frequencies exceeding 10 mHz. A broad enhancement in power dominates the spectrum and we demonstrate it is accurately reproduced using a physics-based model. The results corroborate the scenario where the corona is dominated by Alfv\'enic waves excited in the photosphere by horizontal convective motions, with low-frequency waves subject to reflection at the transition region and higher frequency waves significantly dissipated by the partially ionized chromosphere. The coronal Alfv\'enic power spectrum also indicates there are contributions from \textit{p}-modes (via mode conversion) and a yet-unknown higher-frequency source. These results provide key insight into how the Sun’s convective motions imprint themselves on the corona and highlight the critical role of partial ionization, reflection, and damping in regulating upward-propagating Alfv\'enic waves. A further implication of this is that reconnection-driven Alfv\'enic waves likely play a smaller role in powering the corona and solar wind than has been suggested by recent studies.
\end{abstract}


\section{Introduction} \label{sec:intro}
There is little doubt that the continual interaction between the Sun's outer convective layer and
the magnetic fields is ultimately responsible for the dynamics and appearance of the solar corona. However, we still lack a complete understanding of the different processes that contribute. Magnetohydrodynamic (MHD) waves are one candidate for transferring energy from the turbulent photospheric motions out into the corona and solar wind. The excitation of MHD waves by the solar granulation was suggested nearly 80 years ago by Alfv\'en \citep{Alfven_1947}, with a number of other works supporting his idea \citep[e.g.,][]{1955ApJ...121..461K,Piddington_1956,Ost1961}. Over the subsequent decades it has been suggested that slow and fast magneto-acoustic modes are likely not responsible for heating the corona as their energy is readily dissipated in the lower atmosphere \citep[e.g.,][]{Cranmer_2017,Van_Doorsselaere_2020b}. Alfv\'enic waves\footnote{{Here the use of Alfv\'enic follows from the characteristics outlined in \citep{GOOetal2012}, i.e., namely Alfv\'enic waves are defined by a high degree of incompressibility, dominance of magnetic tension, and a polarisation that is, predominantly, in the transverse direction to the magnetic field.} } can, in principle,  reach the corona - although they are also subject to various processes that lead to reflection and dissipation of their energy in the lower solar atmosphere \citep[see e.g.,][for a review]{Morton_2023}. There are also still many open questions around how the incessantly evolving chromosphere moderates energy transfer to the corona. 

\medskip

The last two decades has seen various investigations into the presence of Alfv\'enic waves in the Sun's corona. They are found to be ubiquitous, with evidence for kink \citep{TOMetal2007, MCIetal2011} and torsional \citep{Morton_2025b} motions across all coronal structures and all through the solar cycle \citep{MORetal2019, Morton_2025c}. While exact values are still debated, the associated energy flux appears to be on the same order as that required to power the quiet Sun corona and solar wind  \citep[$\sim200$~W~m$^{-2}$, e.g.,][]{Withbroe_1977,Aschwanden_2007}.

It is still not evident how the dynamics of small-scale magnetic flux elements in the photosphere imposes itself on the behaviour of the corona. The role of granulation-driven Alfv\'enic waves is of great importance in many models that self-consistently try to explain the hot corona and accelerated solar wind \citep[e.g.,][]{SUZINU2005,  Matsumoto_2018, Matsumoto_2020,  Shoda_2018} or incorporate the effects of wave turbulence to model coronal conditions for prediction \citep[e.g.,][]{van_der_holst_2014,Reville_2020}. Observations of the photosphere reveal that magnetic flux elements are continuously buffeted by the convection, leading to stochastic horizontal motions \citep{BERetal1998, VANBALLetal1998, NISetal2003, Abramenko_2011, CHITetal2012}. There is also evidence of chromospheric waves driven by the horizontal motions \citep{MORetal2013,STAetal2013,STAetal2014}. Photospheric vortices are also common \citep[e.g.,][]{BONetal2008,2019NatCo..10.3504L} and may also excite torsional Alfv\'en waves \citep[e.g.,][]{Shelyag_2013,2021A&A...649A.121B,Breu_2023}. At present, there is no explicit observational evidence that directly connects the photospheric motions to the wave modes observed in the corona.

\medskip

Beyond the inner corona, the solar wind in known to be highly Alfv\'enic in nature. It appears to be dominated by Alfv\'enic turbulence and large amplitude Alfv\'en waves. The large amplitude Alv\'en waves are sudden reversals of the radial magnetic field and corresponding jumps in the radial velocity \citep[referred to as switchbacks, e.g.,][]{Bale_2021}. The source of the Alfv\'enic waves (both the fluctuations and the switchbacks) in the solar wind is still debated, with arguments for coronal origins and in-situ origins. There is currently observational evidence for widespread, small-scale jet activity through the corona \citep[e.g.,][]{Raouafi_2023,Chitta_2023}, which is though to be due reconnection from emerging magnetic flux interacting with the network \citep[e.g.,][]{Raouafi_2014,Chitta_2023b,Chitta_2025}. Both \cite{Raouafi_2023} and \cite{Chitta_2025} make the suggestion that the reconnection events can generate torsional Alfv\'enic perturbations which later end up as switchbacks, although there is currently scant evidence that these jets contain any rotational component. An alternative view is that the switchbacks arise naturally within a state of Alfv\'enic turbulence. \cite{Shoda_2021b} demonstrate that outwardly propagating Alfv\'enic waves can lead to tangential discontinuities if the fluctuation amplitude becomes large \citep[see also][]{Squire_2020}. 

\medskip

As mentioned, the Alfv\'enic waves journey from the photosphere to the corona is subject to a number of challenges. \cite{2019ApJ...871....3S} provide a detailed model of Alfv\'enic wave propagating through the lower atmosphere, incorporating many aspects of the physics. The model is a description of the propagation of torsional Alfv\'en modes in a single flux tube {representing an isolated network magnetic flux tube}, although the general picture described should be applicable to other Alfv\'enic modes. Within the model are the following key components. The lower atmosphere is in a state of partial ionization which leads to efficient damping of high frequency Alfv\'enic waves  \citep[see the review by][]{Soler_2024}. Due to inhomogenities in the magnetic field and density, perpendicular gradients are also present which lead to phase mixing of the waves \citep{HEYPRI1983} and creating the necessary gradients for Ohmic diffusion (enhanced by electron-neutral collisions) to become meaningful in the lower chromosphere, while in the mid and upper chromosphere direct damping by ion-neutral collisions is more relevant \citep[e.g.,][]{SOLetal2013,soler_2015b}.  

The Alfv\'enic waves are also subject to reflection from longitudinal gradients in the Alfv\'en speed \citep[e.g.,][]{1993A&A...270..304V, CRAVAN2005, Magyar_2019}. While gentle in the photosphere and chromosphere, the gradients in the Alfv\'en speed are steep at the transition region which leads to a substantial barrier for low frequency waves. The impact of these phenomena on the Alfv\'enic waves is represented by the frequency-dependent transmission profile of the wave energy flux, shown in Figure~9 of \cite{2019ApJ...871....3S}. The transmission profile represents the transformation of the spectrum of photospheric Alfv\'enic waves by the time it reaches the corona. Aside from an overall reduction in energy flux by two orders of magnitude, the transmission profile shows a frequency dependent structure, peaking between 1-10~mHz (depending on the photospheric magnetic field strength).

{The \cite{2019ApJ...871....3S} model does not contain all potentially relevant physics for the propagation of Alfv\'enic waves from the photosphere to the corona. For example, it has been shown that MHD waves can be strongly coupled in the lower solar atmosphere, with slow, fast and Alfv\'en modes able to convert between each other depending upon the geometry of the plasma and magnetic field \citep[e.g.,][]{CALGOO2008, CALHAN2011, HANCAL2012, Cally_2021}. Hence, potential sources and sinks of Alfv\'enic waves are missed due to focus on Alfv\'en wave solutions only. The magnetic field model is also somewhat simplified, in that it is an isolated magnetic flux tube embedded in an external plasma with ambient field. It does not account for the complex structure of teh magnetic fields that can occur in and around the magnetic network \citep[e.g.,][]{WIEetal2010}. Hence, there are no magnetic nulls or fans present in the lower atmosphere which can also influence wave propagation \citep[e.g.,][]{Galsgaard_2003, Raboonik_2024}.} 
\medskip

The power spectrum of coronal Alfv\'enic waves has been the subject of several previous investigations \citep{MORetal2015, MORetal2016, MORetal2019}, which demonstrate it is composed of a power law component and an enhancement of power around 4~mHz. The previous analysis focused, in part, on measuring the shape of the coronal power spectrum, using an empirically selected function to describe the power enhancement. The origin of the power enhancement was suggested to be due to the mode conversion of \textit{p}-modes to Alfv\'enic waves \citep{TOMetal2007,CALGOO2008, Hindman_2008, MORetal2019}. Recent observations with the Cryogenic Near Infrared Spectropolarimeter \citep[CRYO-NIRSP;][]{Fehlmann_2023} at DKIST \citep{Rimmele_2020} have demonstrated that it is possible to obtain highly detailed measurements of the coronal power spectrum for Alfv\'enic waves \citep{Morton_2025}. These new measurements have superior signal-to-noise and higher temporal resolution that previous similar data. Combined, these two effects provides better sampling of the power spectrum in frequency space and enables waves with frequencies higher than 10~mHz to be examined. The results from \cite{Morton_2025} appear to show the power enhancement cannot be solely due to the \textit{p}-modes due to its extent in frequency space.

\medskip

Our aim here is to understand the origin of the observed shape of the coronal power spectrum. We utilize a theoretically inspired model based on the work of \cite{2019ApJ...871....3S} to describe the broad power enhancement. We also demonstrate that there are additional, narrower peaks of power present in the data, which are modelled with empirical terms. The results indicate the majority of the power enhancement arises from upwardly-propagating, photospheric-driven Alfv\'enic waves and its structure is determined by the wave reflection and damping that occurs in the lower atmosphere. One of the narrower power enhancements can be associated with the mode conversion of \textit{p}-modes, while the other peak is evidence for as yet unidentified contribution.

\section{Data}
\begin{figure}[!ht]
    \centering
    \includegraphics[scale=0.56]{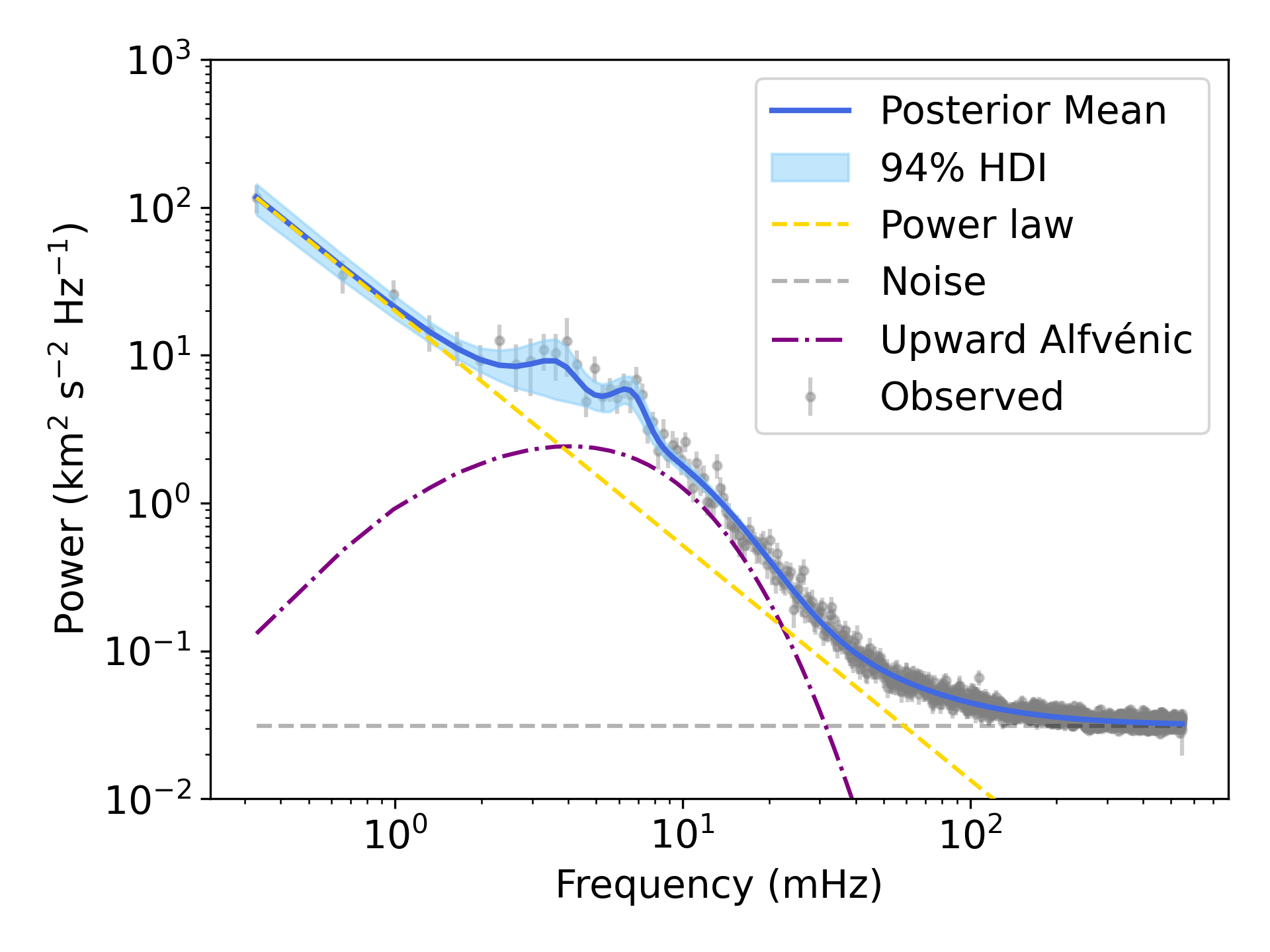}
    \caption{The coronal power spectrum. The measured power spectrum from the Doppler velocities are shown as the grey dots with uncertainties. The posterior mean model of Eq.~\ref{eq:pow_spec} is shown as the solid blue line and the 94\% highest density interval is indicated by the shaded blue region. In addition to this we display the posterior mean estimates for the power law component (yellow dashed), the data noise (grey dashed) and the component of the power from the upwardly-propagating, photospheric-driven Alfv\'enic waves (purple dash-dot).}
    \label{fig:pow}
\end{figure}

We use the Cryo-NIRSP observations taken on the 30 October 2023 that are fully described in \cite{Morton_2025}. The data consists of highly resolved measurements of the Fe XIII 1074~nm infrared coronal line, from which we derive Doppler velocities. \cite{Morton_2025} {demonstrate the Doppler velocities show no correlation with intensity, hence are associated with incompressible fluctuations. Moreover, \cite{Morton_2025b} reveal it is both kink and torsional modes that contribute to the observed signals. Hence, the Doppler velocity fluctuations in this data set are dominated by Alfv\'enic fluctuations}. The observations cover a number of distinct magnetic geometries, but we only use the Doppler velocities from an open field region at a height of 0.1~$R_\odot$ above the limb. Here we do not work with the Doppler velocities directly, but the average power spectrum of the oscillations estimated via a discrete Fourier transform. The measured power spectrum is shown in Figure~\ref{fig:pow} with the uncertainties. We note that the calculated uncertainties appear to be larger than the variability between sequential estimates of power, which could indicate the uncertainties are overestimated.  It was shown in \cite{Morton_2025} that this power spectrum is described by a power law with an excess of power above this. Closer inspection of the data suggests that on top of the enhancement are two narrow peaks at $\sim4$~mHz and $\sim6$~mHz. 

\section{Analysis}

\subsection{Model power spectrum}

In order to model the power spectrum we start with the assumption that there is a power law component and a white noise component, modelling stochastic behaviour and data noise respectively. The origins of the stochastic part of the spectrum is unclear but could potentially arise from reconnection-driven Alfv\'enic waves \citep{Cranmer_2018} or turbulence \citep{Schekochihin_2022}. The model for the enhancement of the power is more involved bringing together previous observational results and theory. It has been suggested by \cite{2019ApJ...871....3S} that the transmission profile at the transition region of Alfv\'enic wave energy flux from upwardly propagating waves excited in the photosphere follows the form
\begin{eqnarray}
        \mathcal{T}(f)&\approx& a_0\frac{1}{\sqrt{2\pi\sigma^2}}\exp\left[-\frac{(log_{10}f-\mu)^2}{2\sigma^2} \right] \nonumber \\
        && \times\left[1+\mbox{erf}\left(\frac{\alpha}{2}\frac{log_{10}f-\mu}{\sigma}\right) \right].
        \label{eq:trans}
\end{eqnarray}
Here the independent variable $f$ is the frequency, $\alpha(B_{phot})$, $a_0(B_{phot})$, $\mu(B_{phot})$, $\sigma(B_{phot})$ are model parameters parameterised by quadratic functions of the photospheric magnetic field strength, $B_{phot}$ \citep[see][ for details]{2019ApJ...871....3S}. In general, the time-averaged energy flux of the Alfv\'enic waves is given by the vertical Poynting flux
\begin{equation}
    \langle S_z\rangle\approx\frac{1}{2}\rho v^2 v_{gr},
\end{equation}
where $\rho$ is the plasma mass density, $v$ is the velocity amplitude, and $v_{gr}$ is the group speed of the wave. The term $v^2$ is equivalent to the power spectrum of the velocity. We note this equation may not be strictly applicable to the corona as there will be a radial dependence of the Poynting flux if we assume that the corona is composed of discrete, isolated flux tubes  \citep{GOOetal2013}. However, there is an on-going debate about whether isolated, discrete structures exist in the corona \citep[e.g.,][]{Malnushenko2022,Kohutova_2024,Uritsky_2024}.

To obtain the expected power spectrum of the coronal waves, the transmission profile in Eq.~(\ref{eq:trans}) needs to combined with the photospheric power spectrum of the Alfv\'enic waves. For this, we use the power spectrum of the horizontal motions of magnetic bright points from \cite{CHITetal2012}. The power spectrum is estimated from the Fourier transform of the auto-correlation function, where the auto-correlation function of photospheric motions is described by a generalized Lorentzian of the form
\begin{eqnarray}
    \mathcal{C}_n(t) = a+\frac{b}{1+\left(\frac{|t_n|}{\tau}\right)^\kappa},
\end{eqnarray}
where $\tau$ is the correlation time, $t_n$ is the delay time (or lag) and $\kappa$, $a$ and $b$ are parameters determined from the observations of the bright point \citep[i.e., here they are considered fixed and we use the values estimated in][]{CHITetal2012}.

We also include two Gaussian functions to describe the peaks around $4$~mHz and 6~mHz, given by
\begin{equation}
    G(f)=D_G\exp\left[-\frac{(f-\mu_G)^2}{2\sigma_G^2} \right],
\end{equation}
where $D_G$, $\mu_G$ and $\sigma_G$ parametrise the Gaussian and are to be determined.

\medskip

Hence the complete model for the coronal power spectrum is given by
\begin{eqnarray}
    S(f)&=&Af^B+C+E\mathcal{T}(f,B_{phot})\mathcal{F}(\mathcal{C}_n(t))+ \nonumber\\
        && G(f, D_{G,1}, \mu_{G,1}, \sigma_{G,1} )+\nonumber\\
        && G(f, D_{G,2}, \mu_{G,2}, \sigma_{G,2}).
        \label{eq:pow_spec}
\end{eqnarray}
We have introduced the parameters $A$, $B$, $C$ and $E$, which are to be determined. The photospheric magnetic field strength, $B_{phot}$, is also left as a free parameter. The function $\mathcal{F}$ indicates the Fourier transform. We note that the Fourier transform of the auto-correlation function suggested by \cite{CHITetal2012} does not provide a correctly normalised power spectral density. From Parseval's theorem, it can be shown easily that
\begin{equation}
    v_{rms}^2=\frac{1}{N}\sum_{i=1}^{N}v_i^2=\frac{1}{N^2}\sum_{k=1}^{N}|V_k|^2,
\end{equation}
where $v_i$ are the values of the velocity time-series and $V_k$ are the Fourier components. We correct the power spectrum such that it provides $v_{rms}=1$~km~s$^{-1}$ \citep[i.e., inline with the direct measurements of the photospheric velocities given in Table~1 of][]{CHITetal2012}.

The parameter $E$ represents the terms in the energy flux that were not accounted for, i.e., density and group speed. It enables the scaling of the transmission profile and photospheric power spectrum to the observed coronal Doppler velocity fluctuations. We do not set the parameter for two reasons: i) we do no know the exact values of $\rho$ and $v_{gr}$ in the photosphere or corona; ii) the coronal Doppler velocity (power) underestimates the true velocity amplitude (power) \citep[see discussions in][]{DEMPAS2012,MORetal2015,Pant_2019,Morton_2025b}.

In order to fit Eq.~\ref{eq:pow_spec} to the data, we employ a Bayesian methodology similar to that discussed in \cite{Morton_2025}. Prior distributions are provided in Appendix~\ref{sec:app_a} for reference.

\begin{figure*}[!ht]
    \centering
    \includegraphics[scale=0.65]{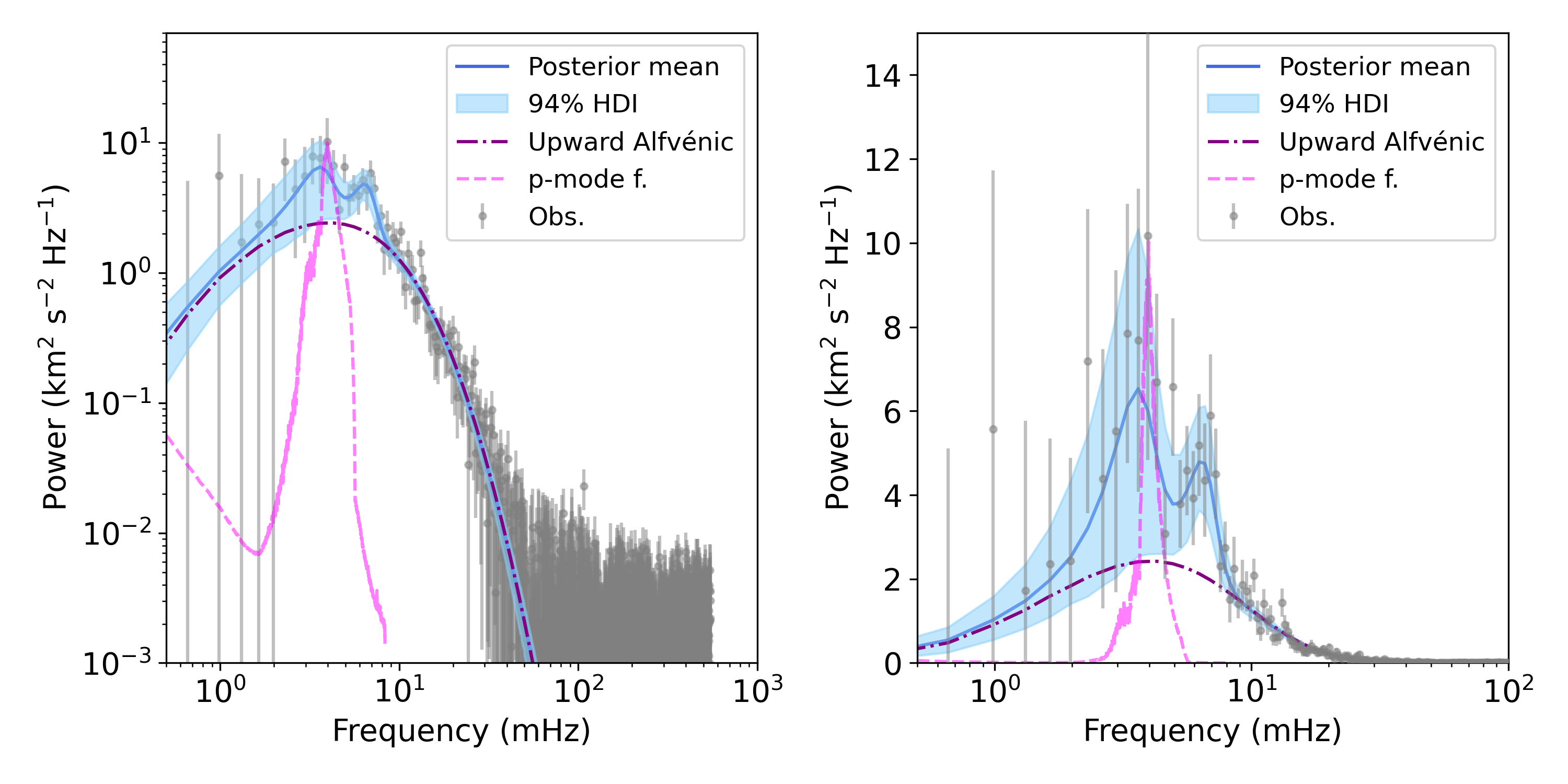}
    \caption{Results of fit to enhancement only. The information is identical in both plots, with the only difference being the scaling of the vertical axes. The measured power spectrum from the Doppler velocities (minus the estimated power law and data noise) are shown as the grey dots with uncertainties. The posterior mean model of the power enhancement from Eq.~\ref{eq:pow_spec} is shown as the solid blue line and the 94\% highest density interval is indicated by the shaded blue region. The contribution of upwardly propagating Alfv\'enic waves excited in the photosphere is indicated by the purple dash-dot curve. The pink dashed curve indicates a coarse estimate for the contribution from mode conversion of \textit{p}-modes to the coronal Alfv\'enic waves.}
    \label{fig:resid}
\end{figure*}

\section{Results}

The results of fitting the proposed model to the measured coronal power spectrum are shown in Figures~\ref{fig:pow} and \ref{fig:resid}. In general, it can be seen that the posterior mean model provides a good description of the data. The 94~\% highest density (credible) interval indicates that in general the model is well constrained by the data, especially at the higher frequencies. The major region of uncertainty coincides with the two narrow peaks that sit atop of the broader power enhancement. In addition to the discussion that follows, we provide a comment on the results for the scaling parameter ($E$) for the photospheric driving in Appendix~\ref{app:scale}.

\subsection{Photospheric-driven waves}
We will now concentrate on the nature of the broad power enhancement. It is perhaps better visualised by removing the power law and data noise components. We do this by subtracting the posterior mean values for the power law and noise components of the model from the observed data. The result is shown in two ways in Figure~\ref{fig:resid}, with the difference only being the log-scaling of the y-axis. The two plots enable visualisation of different aspects of the power enhancement and the estimated model.

In both plots it is evident that the component of the model arising from the upwardly-propagating, photospheric-driven Alfv\'enic waves provides a compelling description of the observed shape of the power enhancement. Given the close correspondence of the data with the proposed model, we can suggest the following interpretation of the observed power spectrum from the physics present in the numerical computation of \cite{2019ApJ...871....3S}. The power enhancement shows an initial rise to a peak around 4-5~mHz (ignoring the narrow peaks for now), which is due to the decreasing reflectivity of waves in the lower atmosphere with increasing frequency. Hence, upwardly propagating Alfv\'enic waves with frequencies smaller than $\sim4$~mHz are substantially reflected in the lower atmosphere. After the peak, the steep decline in power after 10~mHz is largely a signature of the dissipation of Alfv\'enic waves in the chromosphere, arising from the enhanced diffusivity through ion-neutral and electron-neutral collisions. 

The results of \cite{2019ApJ...871....3S} also indicate that the location and spread of (this component of) the power spectrum is ultimately linked to qualities of the magnetic field. The efficiency of the Alfv\'en wave dissipation by ion-neutral collisions in inversely proportional to the Alfv\'en speed, so that the dissipation increases with weaker magnetic fields and decreases with stronger ones. Conversely, the wavelength is directly proportional to the Alfv\'en speed, so that the reflectivity increases with stronger magnetic fields and decreases with weaker ones. The combination of these two effects varies the location of the transmission peak \citep[see Fig.~10 of][]{2019ApJ...871....3S}.   The value of the photospheric field strength estimated from the fit is $B_{phot}=1327\pm90$~G. This value is comfortingly close to the mean values of magnetic field strength obtained for magnetic bright points \citep[$1250-1350$~G, e.g.,][]{Utz_2013}.

\subsection{Other sources of waves}

As mentioned, in addition to the broad power enhancement there are two narrow peaks that sit atop. The peaks are represented in the model by two Gaussian functions, which are empirically chosen, with the central frequencies at $f=3.54\!\mrel{+0.60\\[.12ex]-0.62}$~mHz and $f=6.38\!\mrel{+0.48\\[.12ex]-0.53}$~mHz.  There is less certainty in the posterior about the values for the parameters of these peaks. However, the modelling results suggest that both peaks are genuine, with the 99~\% credible intervals for the amplitudes not containing zero. We note that the tails of the marginal posterior distribution for parameter $D_{G,1}$ are sensitive to the exact choice of prior distribution, while the mean and standard deviation are reasonably robust. This would indicate that the current data cannot constrain the extreme ends of the marginal posterior for $D_{G,1}$ well.

\subsubsection{Origin of the 4~mHz peak}
The peak at 3.54~mHz is consistent with the values found in analysis of the CoMP data \citep{MORetal2019} and is thought to be linked to the linear mode conversion of \textit{p}-modes \citep[e.g.,][]{CALGOO2008,KHOCAL2012}. We believe it is worthwhile trying to compare the observed peak to one expected from the mode conversion process, however it is currently challenging. For the \textit{p}-modes, we use the power spectrum of photospheric Doppler velocity fluctuations described in \cite{1997SoPh..175..287R} and sum over the angular wave number. The waves represented by this power spectrum will be filtered by the acoustic cut-off as they propagate upwards in the lower atmosphere but only below the equipartition layer \citep[][]{Miriyala_2025}. The transformation from fast acoustic to fast magnetic occurs at the equipartion layer, and the fast magnetic waves will not be filtered any further. This happens before the waves reach the region where fast-to-Alfv\'en conversion takes place \citep{CALHAN2011,HANCAL2012}. This means that the coronal Alfv\'enic wave power spectrum arising from mode conversion of the \textit{p}-modes will likely differ from the photospheric spectrum \citep{Miriyala_2025}. Hence, there is an associated transmission profile of the wave energy. 

A crude estimate of this is obtained from the simulation results of \cite{Miriyala_2025}, where we divide the photospheric fast wave power spectrum by the coronal Alfv\'en wave power spectrum in the simulations. A major caveat is that the simulation is of mode conversion in a sunspot atmosphere, so will potentially not be the correct transmission profile for network magnetic fields. Additionally, the simulation is linear and does not include the effects of partial ionisation and diffusivity. However, we use it to provide a coarse estimate of the modification of the \textit{p}-mode power spectrum. The results of multiplying the photospheric Doppler velocity power by the transmission profile from the simulation leads to the pink curves shown in Figure~\ref{fig:resid}. The location in frequency space of the peak is at 4~mHz, which is consistent with the peak location in the coronal power spectrum.

\begin{table}[!t]
    \centering
    \begin{tabular}{l|c|c|c}
      \toprule
      & Mean & 3\% HDI & 97\% HDI \\
      \hline
      \hline
      $A$ & 3.49$\times10^{-4}$ & 3.03$\times10^{-4}$ & 3.94$\times10^{-4}$ \\
      $B$ & 1.59 & 1.54 & 1.63 \\
      $C$ (km$^2$~s$^{-2}$)  & 0.0313 & 0.0311 & 0.0315\\
      $E$ & 3243 & 2480 & 3943\\
      $B_{phot}$ (G) & 1330 & 1180 & 1490\\
      \midrule
      $D_{G,1}$ (km$^2$~s$^{-2}$)  & 4.78 & 0.16 & 8.63 \\
      $\mu_{G,1}$ (mHz)& 3.54 & 2.92 & 4.14\\
      $\sigma_{G,1}$ (mHz) {}& 0.72 & 0.12 & 1.30 \\
      \midrule
      $D_{G,2}$ (km$^2$~s$^{-2}$)  & 2.90 & 1.52 & 4.30 \\
      $\mu_{G,2}$ (mHz) & 6.38 & 5.85 & 6.86 \\
      $\sigma_{G,2}$ (mHz) & 0.85 & 0.41 & 1.32 \\
      \bottomrule
    \end{tabular}
    \caption{Marginal posterior mean and HDI intervals for model parameters.}
    \label{tab:params}
\end{table}

\medskip 

\subsubsection{Origin of the $6$~mHz peak}
The origin of the peak at $f\approx6$~mHz is not immediately evident, at least there is no precedent for such a peak. Hence, we can only speculate on the process responsible. We note that this time-scale has been found to be important in measurements of Alfv\'enic fluctuations in the solar wind \citep{Huang__2024}, with data from the Parker Solar Probe suggesting there is a concentration of energy around $5-8$~mHz. Previous measurements of Alfv\'enic wave power spectra made with the Solar Dynamics Observatory also indicated peaks around 6-8~mHz \citep[see Figure 1 in][]{MORetal2019}, although the genuine nature of the peaks was not able to be confirmed in that analysis.

In \cite{Morton_2025}, a number of options were discussed for the origin of the high-frequency Doppler velocity power associated with the power law component. One of those options would appear to be a suitable candidate for the observed peak, which is Alfv\'enic waves excited in connection with spicules and/or chromospheric/coronal jets. The potential role of spicules in energising the corona has been under intense scrutiny in the past two decades \citep[for a review see, e.g.,][]{2019ARA&A..57..189C}. The spicules are found preferentially along the network boundaries \citep{RUT2006} and the magnetic fields of the network are generally assumed to penetrate into the corona \citep[e.g.,][]{Tuetal2005,WEDetal2009,WIEetal2010}. It is estimated that there are $10^5-10^7$ spicules on the Sun at any one time \citep{Sekse_2012,JUDCAR2010}, with a number density from $0.02-2$~Mm$^{-2}$. Hence they are continually being generated by some underlying mechanism(s). The lifetimes of the spicules is around 400~s \citep{PERetal2014, SKOetal2015}, which is longer than the time-scale associated with the peak.

The relationship between spicules and the range of jets observed in transition region and coronal images (e.g., network jets \citealp{TIAN_2014}, pico-jets \citealp{Chitta_2023}, and jetlets \citealp{Raouafi_2023}) is not clear. However, the transition region/coronal jets are also preferential at network boundaries and estimates for mass flux and kinetic energy flux indicate they have the potential to make a meaning full contribution to energy and mass loss budgets for the solar wind. The lifetimes of such jets is usually $<100$~s, so shorter than the $150$~s time-scale in the coronal power spectrum\footnote{The observed lifetime of the jets may well be an artifact of the constituent plasma being heated and only having emissivity at wavelengths associated with the particular observational bandpass for a reduced period of time. There is precedent for this with spicules \citep{SKOetal2015}.}.  

In general, observations of Alfv\'enic waves in chromospheric spicules and fibrils have typical frequencies are 5-8~mHz \citep[e.g.,][ - although the range is broader, $\sim4-30$~mHz]{JESetal2015}. There are indications that network jets also support Alfv\'enic waves \citep[][although not yet for pico-jets or jetlets]{TIAN_2014}, with comparable amplitudes to the waves in spicules (20~km/s). Some of the wave events observed along the spicules/jets may be associated with photospheric driving due to horizontal motions. There are various other potential sources of waves in spicules/jets although there is no stand-out candidate. For example, \cite{MARetal2017} investigate the generation spicules due to the release of magnetic tension, which excites Alfv\'enic waves with typical frequencies $6-30$~mHz. 

\medskip

Several observations indicate the role of magnetic reconnection in generating spicules \citep[e.g.,][]{He_2009,2019Sci...366..890S} and is known to simultaneously generate waves in numerical simulations \citep[][]{Kigure_2010,Thurgood_2017}. The network and coronal jets are also usually associated with magnetic reconnection of small-scale magnetic fields interacting with the network field \citep{Pontin_2023} despite not having been observed directly. The initial idea of Alfv\'enic waves generated by small-scale reconnection of network fields can be associated with \cite{axford1992}, who suggested high frequency waves ($\geq1$~Hz) would be generated by such events \citep[see also][]{Ruzmaikin_1998}. The timescales of the waves generated will likely depend on the details of the reconnection event \citep[e.g., spatial scales][]{Lynch_2014}. Simulations of reconnection events indicate Alfv\'enic waves with multiple time-scales can be generated. High-frequency ($\sim1$~Hz) waves can be generated in the reconnecting current sheet \citep[e.g.,][]{Yang_2025}. In interchange reconnection, the launch of flux ropes leads to large-amplitude, low frequency waves \citep[long wavelength - set by footpoint separation][]{Lynch_2014, Yang_2025}, or an outflow induced 'whip-like' motion of the ambient field \citep{Yokoyama_1996}. 

\begin{figure}
\centering
    \includegraphics[scale=0.12, trim={15cm 0 15cm 0},clip]{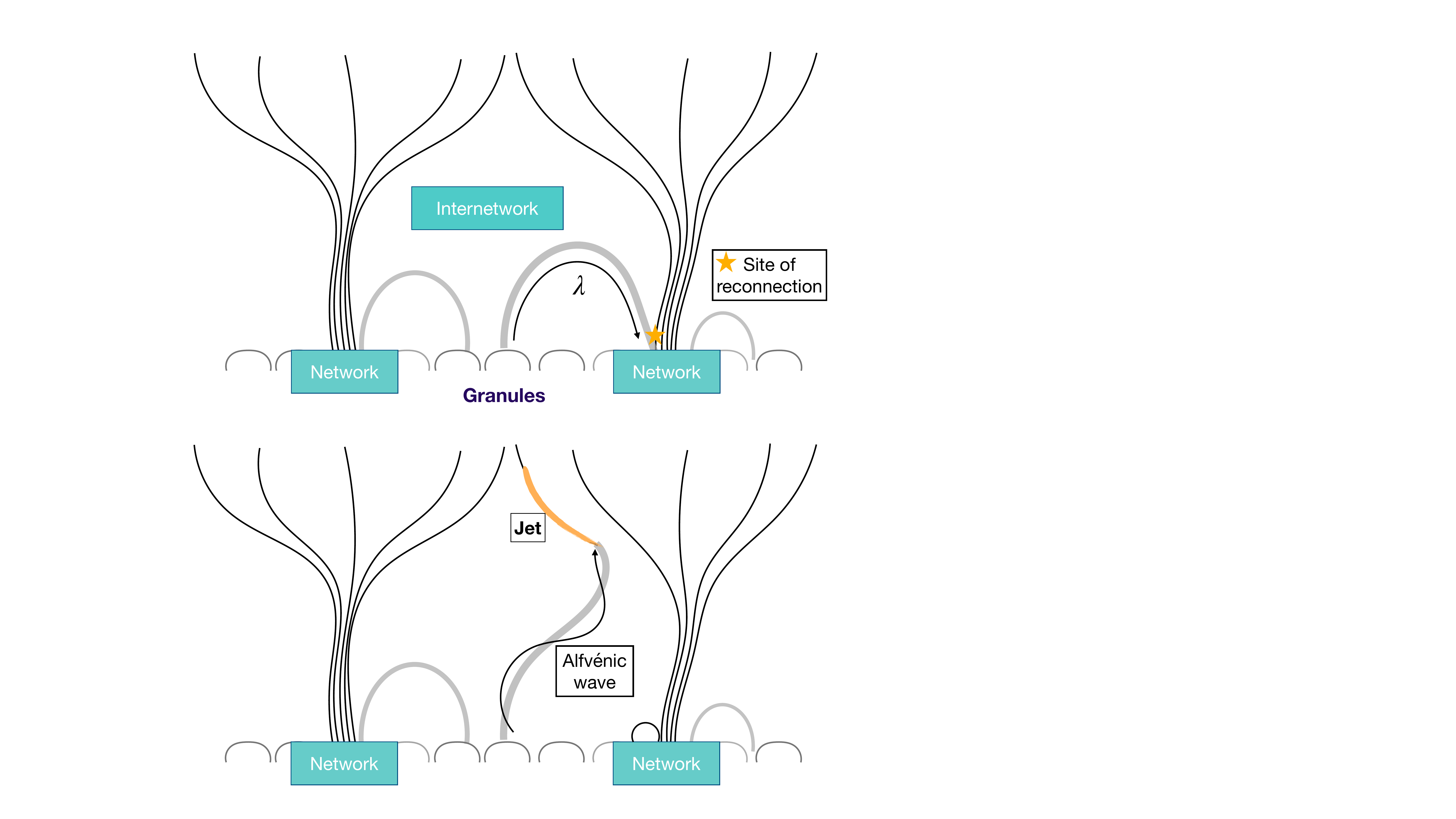}
    \caption{Sketch of Alfv\'enic wave arising from interchange reconnection between chromospheric loops in the internetwork field and the network field.}
    \label{fig:recon_sketch}
\end{figure}

\medskip

Building upon these previous ideas, it is possible to estimate frequencies for waves excited by interchange reconnection based on loop scales. In the lower solar atmosphere, small-scale bipoles appear in the internetwork \citep{GOSetal2014} and at least one of their footpoints move towards the network boundaries. If the polarity is opposite to that of the network field the field may reconnect, leading to an interchange-type reconnection with the network magnetic field. This process leads to the appearance of a jet and the launch of Alfv\'enic waves (sketch in Figure~\ref{fig:recon_sketch}). It was found by \cite{WIEetal2010} the quiescent lower solar atmosphere is dominated by the small-scale loops, with chromospheric loops (reaching heights 500-2500~km) associated with photospheric flux having field strengths $>300$~G. Hence, wavelengths for Alfve\'nic waves will be on the order of the length of these loops, 1.5-8~Mm. Assuming a mass density of $\rho=2.3\times10^{-9}$~kg~m$^{-3}$ \citep[taken from VALC at 1500~km][]{Vernazza_et_al_1981} and a magnetic field strength of around $20$~G in the chromosphere, then one obtains wave frequencies from 5-20~mHz for the Alfv\'enic waves. This range is broadly in agreement with the spicule wave frequencies. The magnetic field strength used here is towards the lower end of estimates at chromospheric heights \citep{Esteban_Pozuelo_2023} and the predicted frequencies increase with an larger values of magnetic field strength (and decreasing mass density).

Alternatively,  mode conversion, different from that suggested by \cite{CALGOO2008}, could also be a source. Recent simulations of non-linear longitudinal-to-transverse mode conversion can occur near the equipartition layer, although this appears to only generate high frequency waves \citep[$\sim$20~mHz][]{Shoda_2018c,Kuniyoshi_2024}. The sensitivity of the current data does not enable us to confirm whether an additional peak occurs at these high frequencies.

It may well be that the $\sim$6~mHz frequency is a signature of the frequency with which wave packets are launched, rather than the intrinsic time scale of a fluctuation. Therefore may represent the envelope of fluctuations with higher frequencies. This would then possibly be linked to the occurrence rate of jets, rather than the fluctuations excited by the spicule generation mechanism. However, from observations of propagating coronal disturbances (interpreted as either mass flows or slow waves) excited by spicules, it appears that the rate at which spicules impact the corona is on the order of 5-15 minutes \citep[e.g.,][]{Samanta_2015}. Although the rate of generation of the propagating disturbances and Alfv\'enic fluctuations need not be related.   

\subsection{Relative contributions}
\begin{figure}
    \centering
    \includegraphics[scale=0.52]{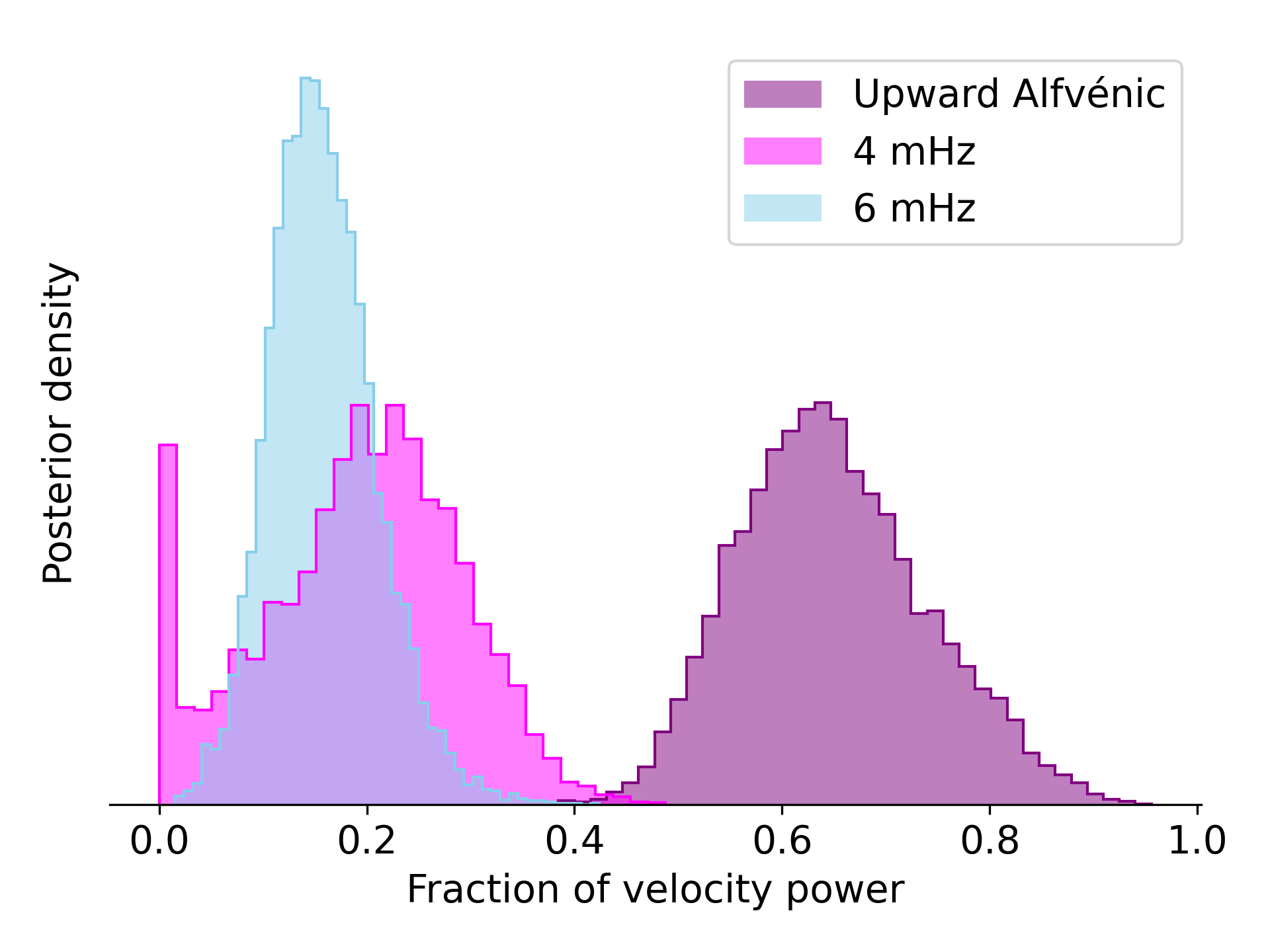}
    \caption{Posterior distributions for the fraction of the coronal Doppler velocity power spectrum provide by each source. The purple corresponds to the photospheric-driven Alfv\'enic waves, fuchsia is the \textit{p}-modes and light blue is the unknown source.}
    \label{fig:perc_source}
\end{figure}

Assuming that the contributions to the power enhancement arises from three distinct sources, it is straightforward to estimate the relative contributions of each source to the coronal Alfv\'enic wave energy flux. Using the posterior draws from our Bayesian model, we calculate the posterior distributions for the percentage of the power enhancement that each source provides (Figure~\ref{fig:perc_source}).

It is estimated that the component related to the upwardly propagating Alfv\'enic waves contributes to $65\pm9$~\% (posterior mean and standard deviation) of the power enhancement, the peak at $\sim4$~mHz
contributes $20\pm9$~\% and the peak at $\sim6$~mHz contributes $15\pm5$~\%.

\medskip

These values indicate that the largest source of Alfv\'enic waves in the corona is from waves excited by convection at the photosphere. The mode conversion of \textit{p}-modes is the next largest source, followed by the unknown contribution. However, the contributions of the \textit{p}-modes and unknown source are rather poorly constrained by the current data

\section{Conclusion}
The new observations provided by DKIST/Cryo-NIRSP provide an unprecedented opportunity to probe the physics of the corona by providing high quality measurements of infrared coronal spectral lines at high spatial and temporal resolution. The Doppler velocities estimated from the spectral lines reveal the corona is rife with incompressible fluctuations that can be associated with Alfv\'enic waves \citep{Morton_2025, Morton_2025b}. Here, we fit a physics-inspired model to the power spectrum of the Doppler velocities in order to identify the origin of the Alfv\'enic waves. We find excellent agreement between the model of \cite{2019ApJ...871....3S} and the coronal Doppler velocity power spectrum, highlighting that the coronal Alfv\'enic waves largely originate from the horizontal motions in the photosphere and the important roles of reflection and chromospheric wave damping in the formation of the coronal power spectrum. This confirms a nearly 80 year old suggestion by \cite{Alfven_1947} and one that lies at the heart of many theories of Alfv\'enic wave propagation through the solar atmosphere. We highlight that this interpretation of the power enhancement contravenes previous suggestions that its presence arose primarily from the mode conversion of \textit{p}-modes \citep[e.g.,][]{TOMetal2007, MORetal2016, MORetal2019}.

{We caveat the above with the recognition that the model of \cite{2019ApJ...871....3S} does not contain a number of potential source and sinks of Alfv\'enic waves in the lower atmosphere \citep[e.g.,][]{Cally_2021}. And, from consideration of Figure~\ref{fig:resid} it is evident that the transmission profile from \cite{2019ApJ...871....3S} does not capture all aspects of the coronal power spectrum.}

In particular, we also demonstrate that there is finer structure in the power spectrum, with two additional narrow peaks. One of these peaks (at 4~mHz) could be associated with the mode conversion of \textit{p}-modes. The results indicate that mode conversion could still make a significant contribution to the flux of coronal Alfv\'enic waves. There also appears to be another unknown source, potentially associated with spicules and/or jetting activity. Estimates for wave frequencies associated with small-scale, interchange reconnection between chromospheric internetwork loops and the network fields are comparable but are subject to various assumptions.

\medskip

Analysis of the total power associated with each of these three sources indicates that the horizontal motions of the photosphere are the biggest contributor to the Poynting flux carried by Alfv\'enic waves in the corona, supplying around 65~\% of the energy. The other two sources contribute the remaining 35~\%. This result seemingly challenges an emerging view that coronal jets provide a dominant component of the Alfv\'enicity of both slow and fast wind streams \cite[e.g.,][]{Raouafi_2023,Chitta_2023,Chitta_2025}. The proportion of coronal velocity power potentially originating from small-scale reconnection would provide less than 20~\% of the total wave energy in the corona. If reconnection driven waves have higher frequencies than 6~mHz, then \cite{Morton_2025} demonstrate that the high-frequency wave power is significantly less than the low-frequency contribution. This is not say that the jets do not make a significant contribution, as they can inject mass and energy (in other forms besides electromagnetic) into the solar wind. 

The role of the granulation as a dominant energy source also ties in with the results from measurements of the coronal perpendicular correlation lengths associated with the fluctuations \citep{Sharma_2023}. The measured coronal correlation lengths are on the order of $L_\perp\sim$8~Mm \citep[see, also][]{Morton_2025,Tajfirouze_2025}, which when traced back to the photosphere corresponds to correlation scales of 1.2~Mm (consistent with scales of granules). Furthermore, there is direct correspondence between the scales associated with coronal Alfv\'enic waves and the relative perpendicular scales of switchbacks when mapped back to the Sun \citep[$3-5^{\circ}$ or $7-13$~Mm;][]{Bale_2021,Fargette_2021}.

\medskip
As a final note, we highlight that the physical picture of wave propagation we outline here is likely relevant for the quiescent regions of the corona (which includes open field regions) that host propagating Alfv\'enic waves. The picture may not be valid for wave behaviour in actives regions. Active region loops show the presence of decay-less kink oscillations \citep[e.g.,][]{TIAetal2012, NISetal2013}, not found in the quiescent Sun. There is evidence that these may be driven by quasi-steady flows \citep{NAKetal2016, KARVAND2020, Zhong_2023}.

\section{Acknowledgments}
RJM is supported by a UKRI Future Leader Fellowship (RiPSAW—MR/T019891/1). Data used has been provided courtesy of NSF and NSO. It is freely available at \url{https://nso.edu/dkist/data-center/}. For the purpose of open access, the author(s) has applied a Creative Commons Attribution (CC BY) licence to any Author Accepted Manuscript version arising. This publication is part of the R+D+i project PID2023-147708NB-I00, funded by MCIN/AEI/10.13039/501100011033 and by FEDER, EU.

The research reported herein is based in part on data collected with the Daniel K. Inouye Solar Telescope (DKIST), a facility of the National Solar Observatory (NSO). NSO is managed by the Association of Universities for Research in Astronomy, Inc., and is funded by the National Science Foundation. DKIST is located on land of spiritual and cultural significance to Native Hawaiian people. The use of this important site to further scientific knowledge is done so with appreciation and respect.

\section{Software}
{Data analysis has been undertaken with the help of NumPy \citep{Numpy}, 
matplotlib \citep{Matplotlib}, IPython \citep{IPython}, Sunpy \citep{sunpy}, Astropy \citep{astropy}, SciPy \citep{Scipy}, PyMC \citep{pymc}.}

\bibliographystyle{aasjournal}


\appendix

\section{Notes on Bayesian model}\label{sec:app_a}

The Bayesian model used to fit the power spectrum is similar to that described in \cite{Morton_2025}. It comprises of a Normal likelihood function for the data, with priors on the model parameters given by:
\begin{eqnarray}
    A &=& \mathcal{N}(5.8\times^{-4}, 2.9\times^{-4}) \\
    B&=& Half Normal(1)\\
    \ln(C) &=& \mathcal{N}(-3.4, 0.5)\\
    \log_{10}(D_{G,1}) &=& \mathcal{N}(0.8, 1)\\
    \mu_{G,1} &=& \mathcal{N}(3,0.5) \\
    \sigma_{G,1}  &=& Half Normal(0.5)\\
    \log_{10}(D_{G,2})  &=& \mathcal{N}(0.8, 1)\\
    \mu_{G,2}  &=& \mathcal{N}(6.4,0.5) \\
    \sigma_{G,2}  &=& Half Normal(0.5)\\
    \log_{10} E &=& \mathcal{N}(3, 1)\\
    \log_{10} B_{phot} &=& \mathcal{N}(3.1, 0.1).
\end{eqnarray}
These priors were formulated based on prior knowledge of coronal velocity power spectrum \citep[e.g.,][]{MORetal2019} and also chosen to provide a prior predictive distribution that does not overly restrict the space of models \citep{Gabryetal2019}. The model was fit using a gradient-based Hamiltonian Monte Carlo sampler \citep{pymc}. 

\section{The scaling term}\label{app:scale}
The model used to fit the data (given in Eq.~\ref{eq:pow_spec}) contains a free parameter $E$ that essentially represents a scaling constant. The nature of $E$ can be seen from the following. The transmission profile, as defined in \cite{2019ApJ...871....3S}, measures the ratio of photospheric Alfv\'enic wave flux to the coronal Alfv\'enic wave flux, i.e.,
\begin{equation}
    \mathcal{T}=\frac{\rho_cv_c^2v_{gr,c}}{\rho_{p}v_{p}^2v_{gr,p}},
\end{equation}
where the subscripts $p$ and $c$ refer to the photosphere and corona, respectively. This expression can be rearranged to describe the coronal velocity power spectrum:
\begin{equation}
    v_c^2 = \frac{\rho_{p}v_{gr,p}}{\rho_cv_{gr,c}}v_{p}^2\mathcal{T}.
\end{equation}
The term $v_{p}^2\mathcal{T}$ is the combination of the transmission profile with photospheric power spectrum present in Eq.~(\ref{eq:pow_spec}). Hence $E$ represents the ratio of density and group speeds in the photosphere and corona. This equation assumes we are measuring the waves as they enter the corona. However, the data we use is taken at a height of $0.1$~$R_\odot$ above the limb, so there will likely be amplification of the waves from the coronal base to the location of measurement due to expected the decrease in density with height \citep[e.g.,][]{MORetal2015}. From a general semi-empirical model of the density profile in the corona \citep{Avrett2008}, the amplification would be a factor $A_1\lesssim1.5$. Additionally, we don't expect much wave damping to occur between the coronal base and the measurement location in the open field region under study as the plasma inhomogeneity is likely small \citep[e.g.,][]{Morton_2021,MorCun2023} and the dissipation mechanisms in the coronal plasma are negligible.

Assuming that $v_{gr}\propto B/\sqrt{\rho}$ (which is reasonable approximation to the kink speed in an open field region where $\rho_e\lesssim \rho_i$),
\begin{equation}
    v_c^2 = \frac{\rho_{p}^{1/2}B_{p}}{\rho_c^{1/2}B_{c}}v_{p}^2\mathcal{T}.
\end{equation}

\noindent We also need to account for the fact that the measured Doppler velocity is lower than the true coronal wave amplitude due to line-of-sight integration effects \citep[e.g.,][]{DEMPAS2012,Pant_2019}. Hence, the term $v_c$ can be expressed as $v_c^2=\frac{A^2_{LOS}}{A_1^2}v_{Dopp}^2$, where $A_{LOS}$ is the factor required to correct for the line-of-sight integration. From our other work with Cryo-NIRSP, we estimate $A_{LOS}^2\approx3600$ \citep{Morton_2025b}. Hence we have
\begin{equation}
    v_{Dopp}^2 = \frac{\rho_{p}^{1/2}B_{p}A_1^2}{\rho_c^{1/2}B_{c}A_{LOS}^2}v_{p}^2\mathcal{T}.
\end{equation}
Using typical values for the photospheric density, $\rho_{p}=3\times10^{-4}$~kg/m$^3$ and coronal density, $\rho_{c}=8\times10^{-13}$~kg/m$^3$, and magnetic field strength, $B_c=5$~G, we find that 
$$
v_{Dopp}^2 \sim 3090v_{phot}^2\mathcal{T},
$$
which is in reasonable agreement with the posterior values $E$ in Table~\ref{tab:params}. We acknowledge there are many assumptions used in arriving at this value, but the fact it is in broad agreement using conservative values is reassuring. We also note that it is likely photospheric motions would not only excite Alfv\'enic waves, but some energy would go to other (magnetoacoustic) MHD modes. This would also influence the value of $E$.

\end{document}